\documentstyle[prb,twocolumn,aps, epsf]{revtex}
\begin{document}
\twocolumn[\hsize\textwidth\columnwidth\hsize\csname @twocolumnfalse\endcsname

\title{In-Plane Magnetolumnescence of Modulation-Doped GaAs/AlGaAs\\ 
Coupled Double Quantum Wells} 
\author{Yongmin\, Kim$^*$, C. H.\, Perry$^{*,\dag}$, and J. A.\, 
Simmons$^{\ddag}$} \address{$^*$National High Magnetic Field 
Laboratory-Los Alamos National Laboratory\\
Los Alamos, NM 87545\\
$^{\dag}$Department of Physics, Northeastern University, Boston, MA
02115\\ $^{\ddag}$Sandia National Laboratory, Albuquerque, NM 87185}
\date{\today} \maketitle
\begin{abstract}
In-plane magnetic field photoluminescence spectra from a series of 
GaAs/AlGaAs coupled double quantum wells show distinctive doublet 
structures related to the symmetric and antisymmetric states.  The 
magnetic field behavior of the upper transition from the antisymmetric 
state strongly depends on sample mobility.  In lower mobility samples, 
the transition energy shows an $\cal N$-type kink with fields (namely a 
maximum followed by a minimum), whereas higher mobility samples have a 
linear dependence.  The former is due to a homogeneous broadening of 
electron and hole states and the results are in good agreement with 
theoretical calculations.
\end{abstract} 

]\narrowtext

The electrical and optical properties of quantum well heterostructures 
have been the subject of considerable investigation for a number of 
years.  Coupled double quantum well (CDQW) structures, where the two 
wells are separated by a thin barrier, are of interest due to an 
enhanced quantum Stark effect and provide an ideal system for studying 
tunneling dynamics.  \cite{Austin,Chen,Little,Perry1,Perry2} 
Modulation-doping of a CDQW creates two parallel two-dimensional 
electron gas (2DEG) layers.  
\cite{Eisenstein,Simmons1,Lyo,Simmons2,Boebinger} This additional 
electronic degree of freedom in the growth ({\bf z}) direction can be 
controlled by varying the barrier thickness, external gate voltages, 
and magnetic fields.  For a strong tunneling structure, wherein two 
quantum wells are separated by a very thin barrier (typically $<$ 
40$\AA$), a tunnel split bandgap ($\Delta$$_{SAS}$) is formed such 
that the symmetric and antisymmetric states are nearly degenerate in 
energy.  For electronic applications, a gated CDQW has a function 
similar to a double barrier resonant tunneling device (RTD).  However, 
an RTD has a side gate structure, whereas a gated CDQW device is an 
entirely planar configuration which has many advantages.  Moon et.  
al.  recently demonstrated that these double layer systems could be 
used as compact digital logic devices constituting a great reduction 
in circuit complexity due to their planar geometry.  \cite{Moon}

In the presence of magnetic fields parallel to the growth direction 
(B$\perp$), the tunnel-split band gap ($\Delta$$_{SAS}$) collapses and 
the structure acts as a single quantum well due to Coulomb 
interactions at certain odd integer Landau filling factors.  
\cite{Boebinger} By contrast, in a purely in-plane field, single 
particle dynamics dominate the interactions between the 2DEGs and 
there is a linear shift in the canonical momentum $\hbar${\bf k} in 
one quantum well with respect to another.  The primary effect is to 
produce a partial energy gap and a strong modulation of the in-plane 
conductance due to an anti-crossing of the {\it k}-dispersion curves 
as demonstrated by Simmons {\it et.  al.}.  \cite {Simmons1,Simmons2}

In a recent theoretical investigation, Huang and Lyo \cite{Huang} 
describe many different phenomena related to the in-plane magnetic 
field dependence of the photoluminescence (PL) spectra.  Their 
calculations show that, as the magnetic field increases, transition 
energy between the antisymmetric and symmetric states displays an $\cal 
N$-type kink, namely a maximum followed by a minimum.  In the presence 
of in-plane magnetic fields the degenerate {\it k}-dispersion curves 
at zero field split in the {\it y}-direction, as depicted in Fig.  1.  
In this report, we present the in-plane magnetic field dependence of 
the PL spectra from a series of CDQWs.  Due to the construction of 
these CDQWs, they exhibit a valence band structure where the 
antisymmetric heavy hole (hh2) state is located between the first 
symmetric heavy-hole (hh1) state and the first symmetric light-hole 
(lh1) state.  The anticrossing of the two dispersion curves in the 
presence of in-plane magnetic fields can produce a partial energy gap.  
\cite {Simmons1,Lyo,Simmons2}This yields large in-plane B-field 
tunable distortions in the Fermi surface and the density of states 
(DOS).  Large maxima and minima are observed in the in-plane 
conductance.  Although we were unable to conclusively observe these 
field tunable effects detected from transport experiments, we did 
observe the $\cal N$-type kink transitions predicted by Huang and Lyo.  
\cite{Huang}

The structures used in this study were modulation doped CDQW 
structures (A, B, and C), each consisting of two GaAs QWs of equal 
width {\it w} separated by an Al$_{0.3}$Ga$_{0.7}$As barrier of 
thickness {\it t}.  Table 1 lists the values of w and t, the total 
electron density {\it n}, and the mobility $\mu$.  Magneto-PL 
measurements were performed at 2K with a 600 mm diameter fiber optic 
system in a 20 tesla superconducting magnet.  A miniature right angle 
prism was attached to the end of the fiber in a $^{4}$He flow cryostat to 
achieve in-plane field geometry.  The data were recorded using a f/4, 
0.3 m spectrograph equipped with a cooled CCD detector.  The PL 
studies were obtained under the following conditions: The laser 
excitation wavelength was 635nm; the power density on the sample did 
not exceed 0.7 mW/cm$^{2}$; and the spectral resolution was approximately 
0.4 meV. More specific details of the experimental techniques have 
been reported elsewhere. \cite{Perry3}

Measured MPL spectra at 2K are displayed in Fig.  2 for sample A and 
C. Spectra for sample B are not shown as its behavior is similar to 
sample A. At zero magnetic field, a broad asymmetric band to band 
transition is observed.  It consists of two peaks corresponding to the 
symmetric to symmetric and antisymmetric to antisymmetric e-hh 
transitions for the lowest two subband states of the CDQW. These are 
denoted as L (lower, symmetric) and U (upper, antisymmetric).  At zero 
field, for sample A, the L peak is more intense than U peak while for 
sample C, the intensities are interchanged as seen in Fig.  2(b).  For 
sample C, the PL intensity transfers from the L to the U transition 
and the L peak disappears around 5T. For sample A, on the other hand, 
the L peak maintains a stronger intensity over the entire field 
region.  The PL intensity and line-shape are intimately governed by 
the homogeneous broadening of electron and hole states.  The 
homogeneous broadening has been shown by Raichev and Vasko 
\cite{Raichev1,Raichev2} to be complicated issue as it is caused 
mainly by impurity scattering, interface roughness, and 
electron-electron scattering.  Homogeneous broadening of electron and 
hole states in a CDQW system is an important factor as the symmetric 
and antisymmetric levels lie in close proximity due to the tunneling 
gap and the electrons in one level can easily be scattered into the 
other level.  Consequently, this interaction can strongly influence 
the PL spectra in a CDQW system as seen in Fig.  2.

In a two-band CDQW system with large homogeneous broadening factor 
($\Gamma$), it is expected that the upper transition has a larger PL 
intensity than the lower transition due to the superposition of the 
upper transition and the tail section of lower transition.  However, 
for a small $\Gamma$ two-band system, one can expect to observe a PL 
spectrum where the lower transition has a higher PL intensity than the 
upper transition because interaction between two states is relatively 
small and hence the two energy levels are well defined.  This is what 
exactly is observed at the zero magnetic field as seen in Fig.  2.  
Sample A has higher mobility than sample C, L peak of sample A (sample 
C) has higher (lower) intensity than U (L) peak such that 
$\Gamma$$_{A}$$<$ $\Gamma$$_{B}$.  

Peak transition energy vs.  magnetic field plots are displayed in Fig.  
3.  For sample A in Fig.  3a, the U and L peaks display approximately 
parallel diamagnetic shift with magnetic field and disappears around 
11T. However, sample C shows completely different behavior (Fig.  3b).  
Initially, the U and L peaks have parallel diamagnetic shift to 4T. 
Beyond 4T, the U peak shows red-shift toward the L peak, namely $\cal 
N$-type transition, a maximum followed by a minimum.  As suggested by 
Huang and Lyo \cite{Huang} who showed that homogeneous broadening 
plays an important role in the behavior of the transition energies in 
high magnetic fields.  These authors clearly show that the in-plane 
magnetoluminescence energy transitions of CDQWs change from parallel 
transitions to $\cal N$-type transitions, as $\Gamma$ varies from 
smaller to larger values.  Our experimental results for sample C are 
in good agreement with their theoretical calculations for a 
$\Gamma$=1.0meV. We see that for a lower mobility (large $\Gamma$) 
system (sample C), due to its large homogeneous broadening of electron 
and hole states, the upper transition smeared in to the lower 
transition at $\sim$5T. However, for a higher mobility (small $\Gamma$) 
system (sample A), due to less homogeneous broadening, the interaction 
between the U and L transitions is relatively small and they show a 
parallel behavior with respect to increasing in-plane magnetic fields.

The integrated intensity for the U and L peaks vs.  magnetic field for 
samples A and C are displayed in Fig.  4.  The U and L peak 
intensities for sample C with respect to the in-plane magnetic field 
shows a behavior comparable to the theoretical calculations in Ref.  
12 for a sample with large $\Gamma$. In the theoretical calculations, the U 
peak decreases its intensity while the intensity of the L peak 
increases with increasing magnetic field.  However, in our experiments 
for sample C, before U peak disappears, U and L peaks show oscillatory 
behavior for all the samples used in this study.  Similar oscillatory 
behavior was observed from sample A and B. We speculate that this 
behavior might be related with non-linear variation of oscillator 
strength in the upper and lower branch while the degenerated 
{\it k}-dispersion curves separate in {\it y}-direction in the presence of 
in-plane magnetic fields.  

In summary, we have measured the PL doublet structure from a series of 
modulation doped CDQWs in the presence of in-plane magnetic fields.  
For transition intensity analysis, we found that initially the upper 
transition is more intense than the lower transition at low fields for 
a low mobility sample.  As the field increases, the electrons in upper 
band depopulate and as a consequence the upper transition merged into 
the lower branch, namely an $\cal N$-type kink.  However, for a high mobility 
sample, the upper and lower transitions show parallel behavior in 
magnetic fields and the upper branch disappears in high magnetic 
fields near 11T. Recent theoretical study which is in good agreement 
with our experimental results reveals that homogeneous broadening of 
electron and hole states are important to understand the transitions 
in the presence of in-plane magnetic fields.

The authors would like to thank S. K. Lyo for helpful discussions and 
suggestions.  Work at NHMFL-LANL is supported through the NSF 
Cooperative Agreement No.  DMR 9527820, the State of Florida, and the 
U.S. Department of Energy.  Work at Sandia National Laboratories is 
supported in part by the Division of Basic Energy Sciences, U.S. 
Department of Energy, No.  DE-AC04-76P00789.

\begin{table}
\caption{Sample parameters of a series of CDQWs.  {\it w}, {\it t}, 
{\it n} and $\mu$ 
represent well width, barrier thickness, 2DEG density and mobility of 
the samples, respectively.  The estimated 2DEG densities are with 
(without) laser illumination conditions.}
\begin{tabular}{lccccc}
&{\it w}/{\it t}&{\it n}&$\mu$ \\ 
&($\AA$)&($10^{11}/cm^{2}$)&($10^{5}cm^{2}$/Vs)\\ \tableline
A(G1176) &150/25&4.8(2.9) &2.7 \\
B(G1177) &100/35&4.2 (2.4) &1.2 \\
C(G1165) &150/15&2.8 (1.6) &0.6 \\
\end{tabular}
\end{table}

\begin{figure}[tb]
\caption{Schematic diagram for the optical transitions from a 
coupled double quantum wells.  (a) At zero field, the transition 
occurs from the antisymmetric and symmetric states denoted as U and L, 
respectively.  (b) In the presence of an in-plane magnetic field 
(B$||$x), the degenerate dispersion curves split and transition occurs 
only at k$\sim$0 at high fields.  }
    \end{figure}
 \begin{figure}[tb]
 \caption{In-plane magnetic field dependent of the PL spectra.  Both 
samples show doublet structures at zero field.  (a) For sample A (high 
mobility), in the entire field region, the lower (L) transition has 
stronger intensity than the upper (U) transition.  (b) For sample C 
(low mobility), the U peak has stronger intensity at zero field but 
transfers to L peak around 4T. The different intensity behaviors are 
due to the homogeneous broadening of electron and hole states (see 
text).}
    \end{figure}

     \begin{figure}[tb]
 \caption{Transition energy vs.  magnetic field plot.  (a) For sample 
A, the U peak runs in parallel to the L peak and disappears around 
11T. (b) Sample C shows $\cal N$-type kink transition, namely a maximum 
followed by a minimum around 4T for the U peak.  }
    \end{figure}

   \begin{figure}[tb]
 \caption{PL intensity vs.  magnetic field plot for sample A (a) and 
 sample C (b).  Both U and L peaks show oscillatory behavior before U 
 peak disappears.  This behavior might be related with non-linear 
 distribution of oscillator strength in the upper and lower branches 
 while the degenerated k-dispersion curves separate in y-direction in 
 the presence of in-plane magnetic fields.  }
    \end{figure}

\end{document}